\documentclass[preprint,showpacs,amsmath,amssymb]{revtex4}

\usepackage{amssymb,amsfonts,graphicx}
\usepackage{dcolumn}
\usepackage{bm}

\begin{document}

\title{Hurst exponent and prediction based on weak-form efficient market hypothesis of stock markets}

\author{Cheoljun Eom}
\affiliation{Division of Business Administration, Pusan National University, Busan 609-735, Republic of Korea}

\author{Sunghoon Choi}
\affiliation{Division of Business Administration, Pusan National University, Busan 609-735, Republic of Korea}

\author{Gabjin Oh}
\affiliation{NCSL, Department of Physics, Pohang University of Science and Technology, Pohang, Gyeongbuk, 790-784, Republic of Korea \& Asia Pacific Center for Theoretical Physics, Pohang, Gyeongbuk, 790-784, Republic of Korea}

\author{Woo-Sung Jung}
\affiliation{Center for Polymer Studies and Department of Physics, Boston University, Boston, MA 02215, USA}

\date{\today}

\begin{abstract}

We empirically investigated the relationships between the degree of efficiency and the predictability in financial time-series data. The Hurst exponent was used as the measurement of the degree of efficiency, and the hit rate calculated from the nearest-neighbor prediction method was used for the prediction of the directions of future price changes. We used 60 market indexes of various countries. We empirically discovered that the relationship between the degree of efficiency (the Hurst exponent) and the predictability (the hit rate) is strongly positive. That is, a market index with a higher Hurst exponent tends to have a higher hit rate. These results suggested that the Hurst exponent is useful for predicting future price changes. Furthermore, we also discovered that the Hurst exponent and the hit rate are useful as standards that can distinguish emerging capital markets from mature capital markets.
\end{abstract}


\maketitle

\section{Introduction}
\label{sec:1}
In the field of finance, the efficient market hypothesis (EMH) proposed by Fama has had a great influence on theory and practice \cite{reference1}. The EMH is based on whether newly generated information is instantaneously and sufficiently reflected in stock prices. Based on the point of time of generation, information can be classified into three types: historical information, public information, and future (or internal) information. Depending on the reflection of each information in stock prices, the EMH can be divided into three types: weak-form EMH of historical information, semi strong-form EMH of public information and strong-form EMH of future information. Recent research in the field of econophysics examining properties and phenomena of financial time-series through interdisciplinary studies are mostly related to weak-form EMH. The fact that weak-form EMH is supported by financial time-series means that historical information such as similar price change patterns is not useful for predicting future price changes. On the other hand, the fact that weak-form EMH is not supported means that similar price change patterns in the past have information values useful for predicting future price changes.

Of the interdisciplinary studies in the fields of finance and econophysics, research on long-term memory properties have been of particular interest \cite{reference2,reference3,reference4,reference5,reference6,reference7}. Many researchers are interested in this research topic because the results of the existence of long-term memory properties not only serves as negative evidence of weak-form EMH but also is closely related to the predictability of stock prices. The Hurst exponent has been widely used as a method to measure long-term memory properties \cite{reference8,reference9,reference10,reference11,reference12}. This measurement quantifies the degree of persistence of similar price change patterns, and it is closely related to weak-form EMH. Also, there are studies that have proposed that the Hurst exponent could be used as an efficiency measurement of stock markets \cite{reference13,reference14}.

However, it is difficult to find studies that have empirically observed the relationship between the existence of long-term memory properties and the predictability of future prices. To investigate such research, it is necessary to establish a prediction model which can examine the practical relationship with the Hurst exponent, which represents a quantitative measurement of the degree of efficiency. Such a prediction model should be based on the use of similar price change patterns, which are common components between weak-form EMH and the Hurst exponent. Accordingly, we used the nearest-neighbor prediction (NN) method, which uses similar price change patterns of the past for the prediction of future price changes \cite{reference15,reference16,reference17,reference18}. This method first sets the past price changes of a specific time period from the previous date, $t$, of a future trading date, $t+1$, to be predicted as the target price change pattern. Next, it selects a price change pattern of the past period similar to the target price change pattern. Finally, in order to predict the price of the future trading date, it uses the price change information generated by following the date of the selected price change pattern of the past. This NN method is known to be useful for short-term prediction.

Accordingly, we empirically investigated the relationship between the predictability calculated from the NN method and the degree of efficiency calculated from the Hurst exponent in financial time-series. The test hypothesis is that the relationship between the degree of efficiency and the predictability is positive. This is due to the fact that we do not support the weak-form EMH, which means that similar price change patterns appear frequently in financial time-series. As the frequency of similar price change patterns increases, the Hurst exponent has a higher value, $H>0.5$. That is, as the degree of efficiency decreases, the Hurst exponent will become greater. Meanwhile, as the similarity of the price change patterns in financial time-series data increases, the predictability using similar price change patterns will increase. In other words, as the degree of efficiency decreases, the predictability increases. In sum, there will be a positive relationship between the degree of efficiency of the Hurst exponent and the predictability of the NN method. In order to empirically investigate this hypothesis, we used 60 stock market indexes of various countries. We found that there is a strongly positive relationship between the degree of efficiency of the Hurst exponent and the predictability of the NN method. That is, a market index with a lower degree of efficiency (a higher Hurst exponent) has a higher level of predictability. These results were commonly observed regardless of data and period. In addition, we discovered that the degree of efficiency of the Hurst exponent and the predictability of the NN method were useful as deterministic factors that can empirically distinguish emerging capital markets from mature capital markets.

In the next section, we describe the data and methods of the test procedures used in this paper. In section \ref{sec:3}, we present the results obtained according to our established research aims. Finally, we summarize the findings and conclusions of the study.

\section{Data and methods}
\label{sec:2}
\subsection{Data}
\label{sec:2-1}
In this study, we used 60 representative stock market indexes of 56 countries (from Datastream). The data period of market indexes is until June 2007. The market indexes of each country are composed of 4 indexes in Africa, 10 indexes in South and North America, 22 indexes in the Asia-Pacific region, and 24 indexes in Europe. This includes 1. Egypt (starting month, January 1995), 2. Kenya (February 91), 3. Morocco (January 94), 4. South Africa (July 95), 5. Argentina (August 93), 6. Canada (January 90), 7. Chile (January 90), 8. Colombia (July 2001), 9. Mexico (January 90), 10. Peru (January 91), the USA [11. DowJones (January 90), 12. S\&P 500 (January 90), 13. NASDAQ (January 90)], 14. Venezuela (June 93), 15. Australia (June 92), 16. China (February 91), 17. Hongkong (January 90), 18. India (January 92), 19. Indonesia (January 90), 20. Israel (January 90), Japan [21. Nikkei 225 (January 90), 22. TOPIX (January 90)], 23. Jordan (June 96), 24. Malaysia (January 90), 25. Oman (November 96), 26. Pakistan (January 91), 27. Philippines (January 90), 28. Russia (January 96), 29. Singapore (January 90), Korea [30. KOSPI (January 90), 31. KOSPI200 (January 90), 32. KOSDAQ (January 97)], 33. Sri Lanka (January 90), 34. Taiwan (January 90), 35. Thailand (January 90), 36. Turkey (January 90), 37. Austria (January 90), 38. Belgium (January 90), 39. Croatia (January 97), 40. the Czech Republic (October 94), 41. Denmark (January 90), 42. Estonia (April 95), 43. Finland (January 90), 44. France (January 90), 45. Germany (January 90), 46. Greece (January 90), 47. Hungary (January 91), 48. Ireland (January 90), 49. Italy (August 93), 50. Luxembourg (January 99), 51. Netherlands (January 90), 52. Poland (January 95), 53. Portugal (January 90), 54. Romania (January 98), 55. Slovakia (January 94), 56. Slovenia (January 94), 57. Spain (January 90), 58. Sweden (January 90), 59. Switzerland (January 90), and 60. the UK (January 90). Further, the return time series, $R_t$, of the market index used in this paper was calculated by the logarithmic change of the price, $R_t=\ln P_t-\ln P_{t-1}$, where $P_t$ is the stock price on day $t$.

\subsection{Method of Hurst exponent}
\label{sec:2-2}
In order to quantify the degree of long-term memory property, we used the Hurst exponent. In previous works, the DFA method \cite{reference11} was the efficient method for accurately calculating the Hurst exponent \cite{reference19}. Therefore, we used the Hurst exponent calculated by the DFA method as a measurement to quantify the degree of efficiency in a financial time series. The Hurst exponent calculated by the DFA method can be explained by the following steps.

First, after the subtraction of the mean, $\overline{x}$, from the original time series, $x(i)$, one accumulates the series, $y(i)$, as defined by

\begin{equation}
y(i)=\sum^{N}_{i=1} [x(i)-\overline{x}],
\label{eq:1a}
\end{equation}

\noindent where $N$ is the number of the time series. Next, the accumulated time series is divided into a window of the same length, $n$. We estimated the trend lines, $y_n(i)$, by using the ordinary least square (OLS) in each window. We eliminated trends existing within the window by subtracting from the accumulated time series in each window. This process was applied to every window and the fluctuation magnitude, $F(n)$, was defined as

\begin{equation}
F(n)=\sqrt{\frac{1}{N}\sum^{N}_{i=1}[y(i)-y_n (i)]^2}.
\label{eq:1b}
\end{equation}

The process mentioned above was repeated for every scale, $n$, $2n$, $3n$, $\ldots$, $dn$. In addition, we investigated whether the scaling relationship exists for every scale. That is, the scaling relationship is defined by

\begin{equation}
F(n)\approx c \cdot n^H ,
\label{eq:1c}
\end{equation}

\noindent where $c$ is the constant and $H$ is the Hurst exponent. Based on $H=0.5$, if $0\le H <0.5$, the time series will have a short-term memory. If $0.5<H\le 1.0$, it will have a long-term memory.

We used the Hurst exponent as the measurement of the degree of efficiency of a stock market in each country. That is, as the Hurst exponent increases, the persistence of similar patterns in past price changes is high. That means that the pattern of past price changes is useful information when predicting future price change. In addition, in order to estimate and use the robust Hurst exponent regardless of the time variation, we used the average Hurst exponent, $\overline{H_j}=\frac{1}{T-1}\sum^{T-1}_{t=1}H^j_t$. This was estimated repeatedly until June 2006, $t=T-1$, by estimating a period of 60 months and rolling 12 months for a whole period until June 2007, $t=T$. The reason why we omitted the data from July 2006 to June 2007 when calculating the mean Hurst exponent is that we needed an out-of-sample of the 12 months prediction period not included in the estimated period of the NN method.

\subsection{Method of nearest-neighbor prediction}
\label{sec:2-3}
In order to investigate the relationship between degree of efficiency of the Hurst exponent and predictability of future price change, we used the hit rate calculated by the NN method. The process of the NN method can be expressed as follows. In the first step, we reconstructed the pattern series, $V^{m, \tau}_n$, with the embedding dimension, $m$, and the time delay, $\tau$, from the financial time series, $x_1$, $x_2$, $\ldots$, $x_T$, and defined as

\begin{eqnarray}
V^{m,\tau}_n = [x_n, x_{n-\tau}, \ldots , x_{n-(m-1)\tau}],\\
n=(m-1)\tau +1, \ldots, T.
\label{2a}
\end{eqnarray}

Here, the structure of the reconstructed time series can be explained as follows. The length of the reconstructed pattern series, $V^{m, \tau}_n$, is $m$ in a time series of $N$ length, and this creates the $N-m+1$ data points. In order to predict price change at time $t+1$, we have to find price change patterns similar to the target price change pattern $V^{m,\tau}_{\textrm{target}}$ at time $t$ among a $N-m+1$ past price pattern series $V^{m,\tau}_n$. The criteria to select the similar patterns in the price change is the distance between both vectors,

\begin{equation}
D=(V^{m,\tau}_{\textrm{target}}-V^{m,\tau}_n)^2 .
\end{equation}

\noindent That is, if the distance is 0, both price change patterns are completely identical. As the distance increases, it will not have a similar pattern of price change. We chose the $K$ number of similar pattern, $V^{m,\tau}_{n,k}$, $k=1,2,\ldots , K$, which has the smallest distance, among the whole pattern series. To confirm whether the chosen similar patterns of past price changes are true, we selected the chosen similar patterns, $V^{m,\tau}_{n,k(*)}$, $k(*)=1,2,\ldots , K^*$, for the embedding dimension,$m+1$ . The future price change, $x_{n+1}$, can be predicted by the next price change, $F(V^{m,\tau}_{n,k(*)})$, of the chosen patterns of the price change, $K^*$, and defined as

\begin{equation}
x_{n+1}=F(V^{m,\tau}_{n,k(*)}).\label{eq:2c}
\end{equation}

We calculated the hit rate by using the degree of consistence as the prediction power by comparing the direction of actual price change with the direction of predicted price change. In order to employ the instinctive method in the estimation of prediction power, we determined the direction by comparing it to the frequency of direction of past price change patterns. That is, the direction of future price, $x_{n+1}$, is determined by the direction having a higher frequency using the number of direction frequency. In addition, this depends on whether the next day price change of the chosen past patterns in the equation \ref{eq:2c} is in an upward direction or downward direction.

We used the same period described above to calculate the Hurst exponent. That is, the period to reconstruct is 60 months. The prediction period is 12 months of out-of-sample, not including the reconstruction period. As in previous studies, the prediction day is one day. That is, using the first reconstruction period, $t=1$, we predicted the direction (up or down) of price change for the first trading day, $t_p=1$, for the prediction period (12 months). Next, we valuated the consistency between actual price change and that estimated by the NN model. To predict the direction of price change on the second trading day, $t_p=2$, we used the pattern series by removing the farthest trading day (1 day) and including the new trading day (1 day). This process was repeated for the prediction period (12 months, $t_p=1,2,\ldots, T_p$). After the prediction period closes, the prediction power can be calculated by the frequency ratio, $NN^j_{t=1}=FQ_t[T_p^{(*)}]/FQ_t[T_p]$, of the number of consistent direction trading, $T_p^{(*)}$, with the number of trading day, $T_p$, for the prediction period. Next, according to the above process after the moving period (12 months) as the process for the Hurst exponent, we calculated the hit-rate $NN^j_{t=2}$ of future price change using the second reconstruction period, $t=2$. This process was repeated until the reconstruction period, $t=T-1$, of the last day of the prediction period. We used the average hit-rates, $\overline{NN_j}=\frac{1}{T-1}\sum^{T-1}_{t=1}\overline{NN^j_t}$, $t=1,2,\ldots,T$, in terms of the consistent direction between actual price changes and those estimated by the NN model as the prediction power in order to observe the relationship with the degree of efficiency of the Hurst exponent.

\section{Results}
\label{sec:3}
The purpose of this study is to empirically investigate the relationship between the degree of efficiency and the predictability in financial time-series data; that is, the test hypothesis is whether the relationship between the degree of efficiency and the predictability is positive. The Hurst exponent was used as the measurement of the degree of efficiency and the NN method was used for the prediction of the directions of future price changes.

\begin{figure}
\centering
\includegraphics[width=1.0\textwidth]{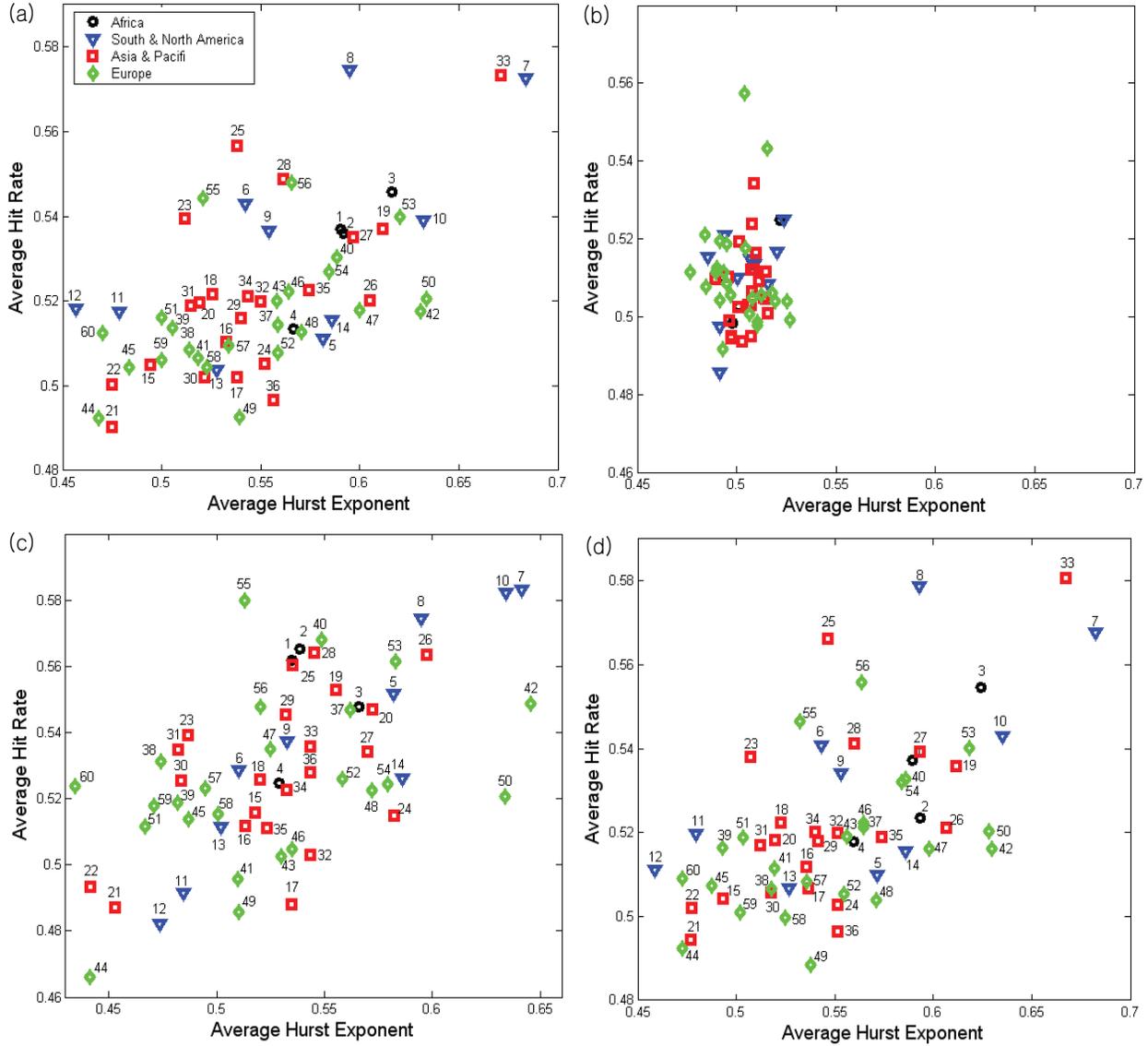}
\caption{This figure shows the results of the empirical investigation of the relationship between the degree of efficiency and the predictability in financial time-series data. The X-axis denotes the average Hurst exponent, while the Y-axis indicates the average hit rate. Fig. \ref{fig:1}(a) represents the relationship between the average Hurst exponent and the average predictability from the estimation period of 60 months, the prediction period of 12 months and the rolling period of 12 months in the whole period until June 2007. Fig. \ref{fig:1}(b) were obtained by using the random time-series data. Fig. \ref{fig:1}(c) shows the results observed using recent data for the past three years, June 2004 to July 2007. Fig. \ref{fig:1}(d) shows the results observed by changing the prediction and rolling periods to 6 months. The market indexes were divided into four types depending on regions. That is, the black circles, blue triangles, red boxes, and green diamonds indicate Africa, North and South America, the Asia-Pacific region, and Europe, respectively.}
\label{fig:1}       
\end{figure}

The results are showed in Fig. \ref{fig:1}. In Fig. \ref{fig:1}, the measurements represent the averages of the values repeatedly measured depending on estimation, prediction and rolling periods in the whole period. Here, the X-axis denotes the average Hurst exponent, $\overline {H_j}$, while the Y-axis indicates the average hit rate, $\overline{NN_j}$. Fig. \ref{fig:1}(a) represents the relationship between the average Hurst exponent and the average hit rate for the estimation period of 60 months, the prediction period of 12 months and the rolling period of 12 months in the whole period (to June 2007). Fig. \ref{fig:1}(b) has the same conditions as Fig. \ref{fig:1}(a). But the difference is that the results of Fig. \ref{fig:1}(b) were obtained by using the time-series data that follow the random walk process generated by adopting the mean and standard deviation of each market index as parameters. Fig. \ref{fig:1}(c) has the same conditions as Fig. \ref{fig:1}(a). But the difference here is that the results were observed using only data for the past three years; that is, June 2004 to July 2007. Fig. \ref{fig:1}(d) shows the results observed for more various periods by changing the prediction and rolling periods to 6 months. The market indexes indicated in each figure were divided into four types depending on regions. That is, the black circles, blue triangles, red boxes, and green diamonds indicate Africa (4 indexes), North and South America (10 indexes), the Asia-Pacific region (22 indexes), and Europe (24 indexes), respectively.

According to the results, we empirically discovered that there is a close relationship between the degree of efficiency and predictability in financial time-series data. In other words, there was a highly positive relationship, $\rho(\overline{H_j},\overline{NN_j})=60.37\%$, between the average Hurst exponent, $\overline{H_j}$, the measurement of the degree of efficiency, and the average hit rate, $\overline{NN_j}$, the measurement to predict the directions of future price changes. Additionally, three results were presented to examine the robustness of these results. First, Fig. \ref{fig:1}(b) represents the results observed using random time-series data. In this figure, we could not find the results observed in Fig. \ref{fig:1}(a). Second, Fig. \ref{fig:1}(c) represents the results using data of the past three years only; that is, June 2004 to June 2007. We also found a positive relationship, $\rho(\overline{H_j},\overline{NN_j})=58.16\%$. Third, Fig. \ref{fig:1}(d) indicates the results using the data of more various market situations by changing prediction periods and rolling periods to 6 months. Here, we also found a strongly positive relationship, $\rho(\overline{H_j},\overline{NN_j})=60.08\%$. These results suggested that, due to the high frequency of similar price change patterns in the past, a market index with a lower degree of efficiency tends to have a higher Hurst exponent and a higher hit rate on the average. That is, the average Hurst exponent has a positive relationship with the average hit rate.

In addition, the areas of figure \ref{fig:1} were divided into two based on the average Hurst exponent and the average hit rate. Afterwards, the location of each market index was investigated. The first position is the top right of the figure. This area has a high average Hurst exponent and a high average hit rate due to a lower degree of efficiency. Countries with market indexes belonging to this area include Chile, Colombia, Peru, Indonesia, Oman, Philippines, Russia, SriLanka, Estonia, Luxembourg, Portugal, and Slovenia. That is, these stock market indexes are those belonging to emerging capital markets. The second position is the bottom left of the figure. This area has a lower average Hurst exponent and a lower average hit rate due to a high degree of efficiency. Countries with market indexes belonging to this area include the USA, Australia, Japan, South Korea, France, Germany, Italy, Switzerland, and the UK. That is, these market indexes are those belonging to mature capital markets. These results suggested that the two measurements, the Hurst exponent and the hit rate, are significant standards which can distinguish emerging capital markets from mature capital markets.

\section{Conclusions}
We empirically investigated the relationship between the degree of efficiency and the predictability in financial time-series data. The Hurst exponent was used as the measurement of the degree of efficiency; in addition, the hit rate calculated from the NN method was used for the prediction of the directions of future price changes. In the test hypothesis established, there is a positive relationship between the degree of efficiency of the Hurst exponent and the predictability of the hit rate calculated by the NN method. In order to empirically examine the test hypothesis, we used 60 market indexes of various countries. In addition, in order to ensure the robustness of our observations, results were conducted through changes of data, analysis periods, and prediction periods. According to the results, we empirically discovered that there is a close relationship between the degree of efficiency and the predictability of the financial time-series data. That is, a market index with a higher Hurst exponent tends to have a higher hit rate on the average. These results suggested that the Hurst exponent represents a measurement which has information values useful for the prediction of future price changes. Furthermore, we also found that the Hurst exponent and the hit rate are useful as standards that can distinguish emerging capital markets from mature capital markets.



\end{document}